\documentclass[reprint,superscriptaddress,aps,twocolumn,showpacs]{revtex4-2}
\usepackage{amssymb,amsmath}
\usepackage{graphicx}
\graphicspath{{figures/}} % Location of the graphics files
\usepackage{braket,mleftright}
\usepackage{xcolor}
\usepackage{bm}
\usepackage[caption=false]{subfig}
\allowdisplaybreaks

\begin{document}
	
\title{Study of $N(1520)$ and $N(1535)$ structures via $\gamma^*p\to N^*$ transitions}
	
\author{A. Kaewsnod}
\email[]{a.kaewsnod@gmail.com}
\author{K. Xu}
\email[]{gxukai1123@gmail.com}
\author{Z. Zhao}
\affiliation{School of Physics and Center of Excellence in High Energy Physics and Astrophysics, Suranaree University of Technology, Nakhon Ratchasima 30000, Thailand}
\author{X. Y. Liu}
\affiliation{School of Physics and Center of Excellence in High Energy Physics and Astrophysics, Suranaree University of Technology, Nakhon Ratchasima 30000, Thailand}
\affiliation{School of Physics, Liaoning University, Shenyang 110036, China}
\author{S. Srisuphaphon}
\affiliation{Department of Physics, Faculty of Science, Burapha University, Chonburi 20131, Thailand}
\author{A. Limphirat}
\author{Y. Yan}
\email[]{yupeng@g.sut.ac.th}
\affiliation{School of Physics and Center of Excellence in High Energy Physics and Astrophysics, Suranaree University of Technology, Nakhon Ratchasima 30000, Thailand}
	
\date{\today}
	
\begin{abstract}
The helicity amplitudes of the $N(1520)$ and $N(1535)$ resonances in the $\gamma^*p\to N^*$ electromagnetic transition are studied in the constituent quark model using the impulse approximation, with the proton and resonances assumed to be in three-quark configurations.
The comparison of theoretical results and experimental data on the helicity amplitudes $A_{1/2}$, $A_{3/2}$, and $S_{1/2}$ indicates that the $N(1520)$ and $N(1535)$ resonances are primarily composed of three-quark $L=1$ states but may contain additional components.
However, it is improbable that contributions from meson clouds will be dominant at low $Q^2$.
%The helicity amplitudes of the $N(1520)$ and $N(1535)$ resonances in the $\gamma^*p\to N^*$ electromagnetic transition are studied in the constituent quark model in the impulse approximation with assuming that the proton and the resonances are in the three-quark configurations. 
%The comparison between the theoretical results and experimental data on the helicity amplitudes $A_{1/2}$, $A_{3/2}$, and $S_{1/2}$ reveals that the $N(1520)$ and $N(1535)$ resonances are mainly three-quark $L=1$ states but may
%consist of other components. However, it is unlikely that meson-cloud contributions may be dominant at low $Q^2$.
		
%\keywords{Permutation group, Spatial wave function, Form factor, Helicity amplitude}
\end{abstract}
	
\maketitle
	
\section{\label{sec:introduction} Introduction}

Low-lying nucleon resonances such as $N(1440)$, $N(1520)$, and $N(1535)$ have been investigated continuously since their discovery, but their inner structure remains an open question.
In the three-quark picture, for example, theoretical works always predict that the $N(1440)$ has a greater mass than the $N(1520)$ and $N(1535)$.
Information about the electromagnetic structure of the nucleon resonances has been accumulated by the experimental facilities like CLAS (JLab), MAID (Mainz) and MIT-Bates.
In recent decades, the $\gamma^*p\to N(1520)/N(1535)$ helicity amplitudes in a wide range of $Q^2$ have been obtained and analyzed at JLab \cite{PhysRevLett.86.1702,PhysRevC.76.015204,PhysRevC.80.015205,PhysRevC.71.015201,PhysRevC.78.045209,Aznauryan:2009mx,PhysRevC.68.065204,PhysRevC.86.035203,Mokeev:2012vsa} and by the Mainz group \cite{Drechsel:2007if,Tiator:2011pw,Tiator:2009mt,Drechsel:2007if} as one parametrization of the inner structures. 

Quark models such as the light-front (LF) relativistic quark model \cite{Capstick:1994ne,PACE200033,PhysRevC.85.055202,PhysRevC.95.065207}, covariant spectator quark model \cite{PhysRevD.95.054008,PhysRevD.95.014003}, and chiral quark models \cite{PhysRevD.84.051301,PhysRevD.85.093014,PhysRevC.77.065207,Golli:2013uha} have been applied to study the helicity amplitudes of $N(1520)$ and $N(1535)$.
The fact that the results from the quark models are comparable with the experimental data of the $\gamma^*p\to N(1520)$ and $\gamma^*p\to N(1535)$ helicity amplitudes at large square momentum transfer $Q^2$ indicates that these two lowest negative-parity nucleon resonances are dominated by the three valance quarks.
However, at low and medium $Q^2$, the poor theoretical results in the quark models may indicate that other components, in addition to the three valance quarks, may play an important role.
The meson cloud effects are expected to be significant for both $N(1520)$ and $N(1535)$ \cite{PhysRevC.85.055202,JuliaDiaz:2007fa,PhysRevD.95.014003,PhysRevD.84.051301,PhysRevD.85.093014,PhysRevC.77.065207}.
The inclusion of pentaquark components $q^4\bar q$ in Refs. \cite{PhysRevC.77.065207,An2009} leads to a reasonable description of the experimental results of $A_{1/2}$ of $N(1535)$.
The mass ordering problem of $N(1440)$, $N(1520)$, and $N(1535)$ is investigated by including the ground state pentaquark components in the negative-parity nucleon resonances \cite{PhysRevD.101.076025}, pointing out that the $N(1520)$ and $N(1535)$ may contain large pentaquark components.

In this work, we study the helicity transition amplitudes of the $N(1520)$ and $N(1535)$ resonances in the process $\gamma^*p\to N^*$ in the three quark picture.
The work is expected to draw the upper limit of the three-quark contribution and to serve as springboard for a more comprehensive study that will include meson clouds, pentaquark components, and meson-baryon molecular components.
The proton wave function in this work is adopted from the study of the proton electric form factor $G_E$ in Ref.\cite{PhysRevD.105.016008}.
This work employs the impulse approximation, with the electromagnetic current being induced by free single quark currents.
The impulse approximation has been verified as a good approximation of real processes.
For instance, the experimental data on the proton's charge form factor and transition $\gamma^*p\to N(1440)$ are well reproduced in the three quark picture using the impulse approximation \cite{PhysRevD.105.016008}.

The paper is organized as follows.
In Sec. \ref{sec:result}, we briefly introduce the helicity amplitude formalism before fitting the $A_{1/2}$, $A_{3/2}$, and $S_{1/2}$ of the $N(1520)$ and $N(1535)$ resonances to the experimental data.
Sec. \ref{sec:Summary} contains a concise summary.
	
\section{\label{sec:result}Helicity amplitudes of $N(1520)$ and $N(1520)$ resonances}
\vspace{-1\baselineskip}
\begin{figure}[h]
	\includegraphics[width=0.3\textwidth]{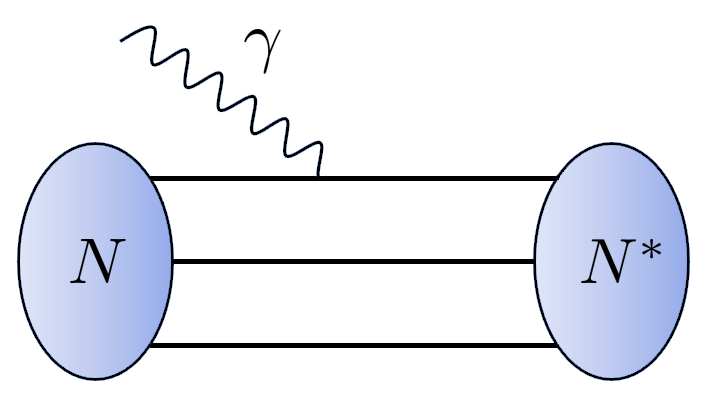}
	\vspace{-1\baselineskip}
	\caption{Diagram of photoproduction transition $N\gamma^*\rightarrow N^*$ where both the initial nucleon ($N$) and final nucleon resonance ($N^*$) are in the three-quark picture.}\label{fig:quarkline}
\end{figure}

We extend the constituent quark model in the work \cite{PhysRevD.105.016008} to investigate the helicity amplitudes of electromagnetic transitions $\gamma^* N\rightarrow N(1520)/N(1535)$, where the electromagnetic current is induced by the free single quark current (the impulse approximation) as illustrated in Fig. \ref{fig:quarkline}.
We assumed that both the nucleon $N$ and the resonance $N^*$ are composed of three quarks.

%The momentum of the nucleon $P_i$, the momentum of the resonance state $P_f$, and the photon momentum $k=P_f-P_i$ are defined as $P_i=(E_N,0,0,-|\bm k|)$, $P_f=(M_{N^*},0,0,0)$, $k=(\omega,0,0,|\bm k|)$.
The square of the four-momentum transfer is expressed as $Q^2=-k^2=|\bm k|^2-\omega^2$ with the photon energy $\omega$ and the magnitude of the photon momentum $|\bm k|$, defined as
\begin{alignat}{3}\label{eq:p photon}
	&\omega&&=\ K-\frac{Q^2}{2M_{N^*}},\nonumber\\
	&|\bm k|&&=\ \left[\omega^2+Q^2\right]^\frac{1}{2}.
\end{alignat}
The $K$ in Eq. (\ref{eq:p photon}) is the real-photon momentum in the $N^*$ rest frame, defined as
\begin{alignat}{3}
K=\frac{M_{N^*}^2-M_N^2}{2M_{N^*}},
\end{alignat}
where $M_{N^*}$ and $M_N$ are respectively the $N^*$ and $N$ masses.

The transverse helicity amplitudes $A_{1/2}$ and $A_{3/2}$, and the longitudinal helicity amplitude $S_{1/2}$ are usually defined in the $N^*$ rest frame as
\begin{alignat}{3}
&A_{i}&&=&&\ \dfrac{1}{\sqrt{2K}}\Braket{N^*,S'_z=i|q_1'q_2'q_3'}\nonumber\\
&&&&&\times T(q_1q_2q_3\rightarrow q_1'q_2'q_3')\Braket{q_1q_2q_3|N,S_z=i-1},\nonumber\\
&S_{1/2}&&=&&\ \dfrac{1}{\sqrt{2K}}\Braket{N^*,S'_z=\frac{1}{2}|q_1'q_2'q_3'}\nonumber\\
&&&&&\times T(q_1q_2q_3\rightarrow q_1'q_2'q_3')\Braket{q_1q_2q_3|N,S_z=\frac{1}{2}},
\label{eq:helicity}
\end{alignat}
where $i=\frac{1}{2}$ for $A_{1/2}$ and $i=\frac{3}{2}$ for $A_{3/2}$.
The $N$ wave function $\Braket{q_1q_2q_3|N,S_z}$ and $N^*$ wave function $\Braket{q_1'q_2'q_3'|N^*,S'_z}$ in the three-quark picture have been described in details in Refs.\cite{PhysRevC.100.065207,PhysRevD.101.076025}, and we refer the readers to those works for additional information.
The $T(q_1q_2q_3\rightarrow q_1'q_2'q_3')$ in Eq. (\ref{eq:helicity}) is the transition amplitude of the process $\gamma q\to q'$ displayed in Fig. \ref{fig:quarkline}, which can be calculated in the standard language of quantum field theory,
\begin{align}\label{eq:Matrix elements}
	T(q_1q_2q_3\rightarrow q_1'q_2'q_3')=\ &e_3\bar u_{s'}(p')\gamma^\mu u_s(p)\epsilon_\mu^\lambda(k)\Braket{q'_1q'_2|q_1q_2}\nonumber\\
	=\ &e_3T_{s's}^\lambda\Braket{q'_1q'_2|q_1q_2}.
\end{align}
Here $e_3$ and $u_{s(s')}$ denote the electric charge and the Dirac spinners of the third quark with the single quark spin projections $s',\,s$, and $\lambda$ is the photon helicity.
In the work, the mass of $u$ and $d$ quarks are taken to be $m=5$ MeV.
The photon polarization vectors $\epsilon_\mu^\lambda(k)$ of the longitudinal ($\lambda=0$) and transverse ($\lambda=1$) helicity amplitudes in the Lorentz gauge are defined as $\epsilon^0_\mu=(1,0,0,0)$ and $\epsilon^+_\mu=-\frac{1}{\sqrt{2}}(0,1,i,0)$.
The transition amplitude in the longitudinal helicity amplitude $S_{1/2}$ is defined by the time component of the electromagnetic current in Refs.\cite{Dreschsel_1992,DRECHSEL1999145,Drechsel2007}.
The matrix elements $T_{s's}^\lambda$ are shown in Appendix \ref{app:T helicity}.
In Eq. (\ref{eq:helicity}), one sums over all possible quantum states of the intermediate three quarks and integrates over the momenta of all quarks.

\begin{figure}[!t]
		\subfloat{%
		\label{fig:A12_N1520}
		\includegraphics[clip,width=0.8\columnwidth]{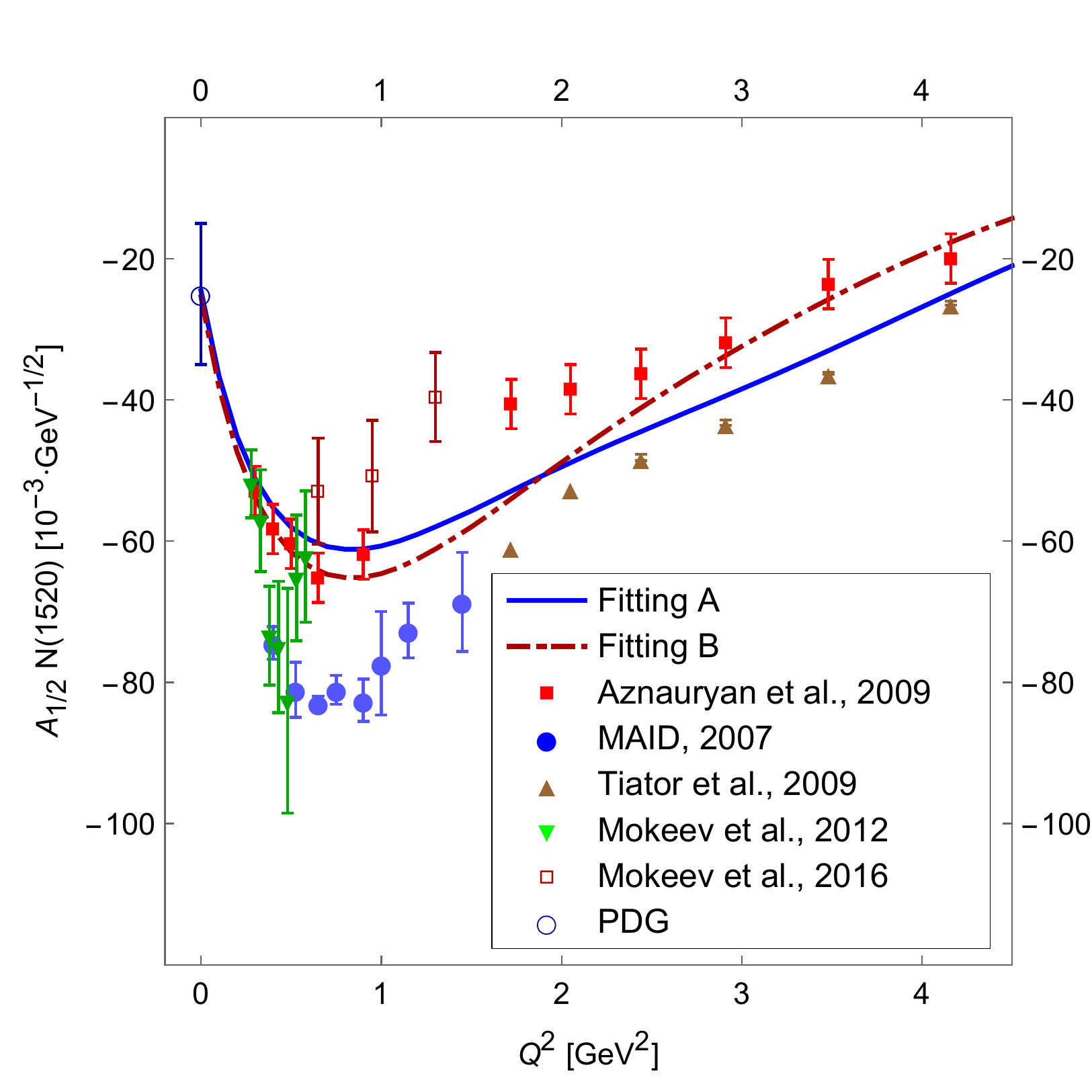}%
	}
\llap{
	\parbox[b]{4.4in}{\textcolor{white}{\large a}\\\rule{0ex}{2.25in}
}}
\llap{
	\parbox[b]{0.77in}{{\scriptsize \cite{Aznauryan:2009mx}}\\\rule{0ex}{0.932in}
}}
\llap{
	\parbox[b]{1.63in}{{\scriptsize \cite{Drechsel:2007if}}\\\rule{0ex}{0.815in}
}}
\llap{
	\parbox[b]{1.2in}{{\scriptsize \cite{Tiator:2009mt}}\\\rule{0ex}{0.698in}
}}
\llap{
	\parbox[b]{1.18in}{{\scriptsize \cite{PhysRevC.86.035203}}\\\rule{0ex}{0.58in}
}}
\llap{
	\parbox[b]{1.25in}{{\scriptsize \cite{PhysRevC.93.025206}}\\\rule{0ex}{0.47in}
}}
\llap{
	\parbox[b]{2.43in}{{\scriptsize \cite{Zyla:2020zbs}}\\\rule{0ex}{0.355in}
}}
	\vspace{-1\baselineskip}
	
	\subfloat{%
		\label{fig:A32_N1520}
		\includegraphics[clip,width=0.8\columnwidth]{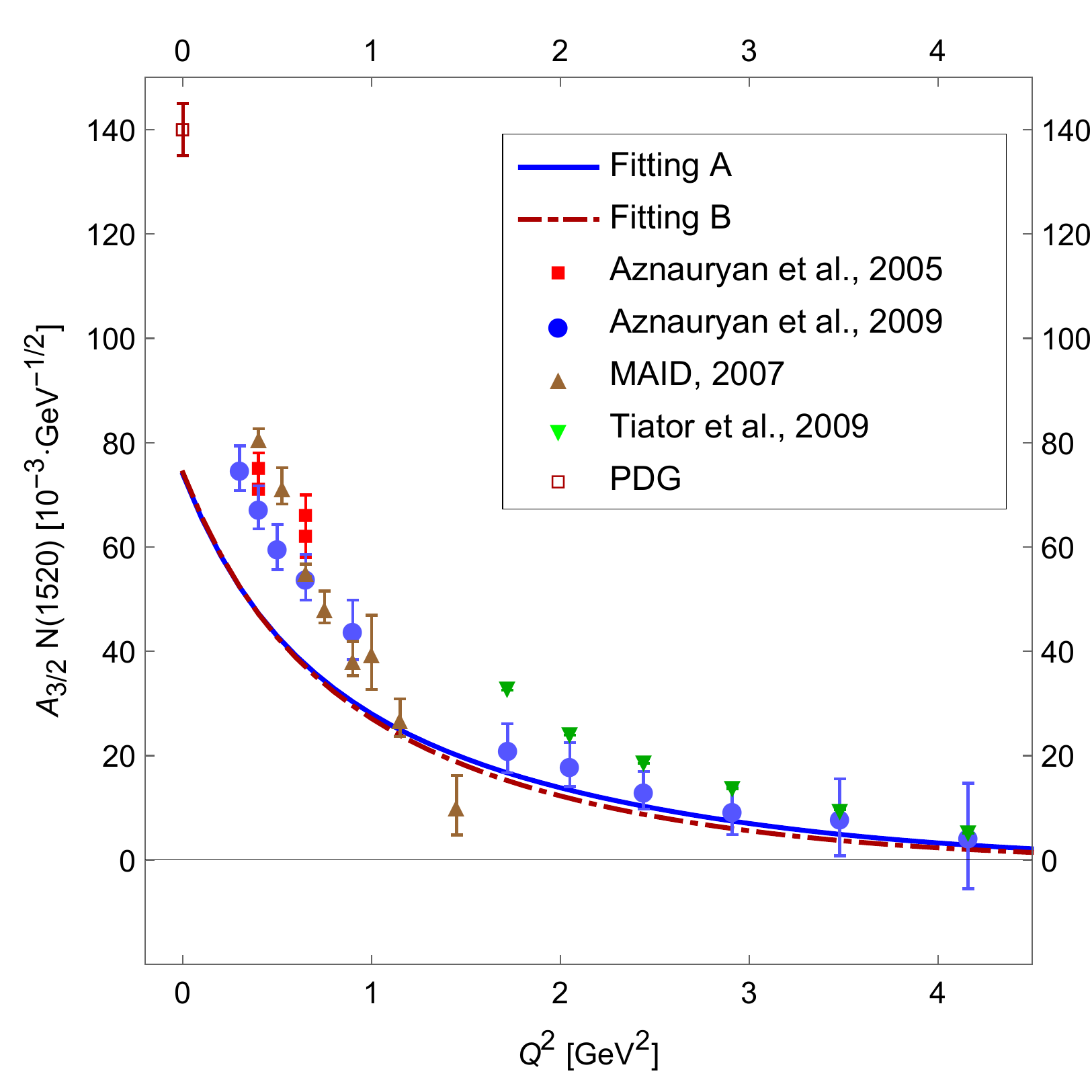}%
	}
	\llap{
	\parbox[b]{4.5in}{\textcolor{white}{\large b}\\\rule{0ex}{2.1in}
}}
\llap{
	\parbox[b]{0.66in}{{\scriptsize \cite{PhysRevC.71.015201}}\\\rule{0ex}{1.986in}
}}
\llap{
	\parbox[b]{0.75in}{{\scriptsize \cite{Aznauryan:2009mx}}\\\rule{0ex}{1.856in}
}}
\llap{
	\parbox[b]{1.65in}{{\scriptsize \cite{Drechsel:2007if}}\\\rule{0ex}{1.726in}
}}
\llap{
	\parbox[b]{1.3in}{{\scriptsize \cite{Tiator:2009mt}}\\\rule{0ex}{1.596in}
}}
\llap{
	\parbox[b]{2.28in}{{\scriptsize \cite{Zyla:2020zbs}}\\\rule{0ex}{1.46in}
}}
	\vspace{-1\baselineskip}
	
	\subfloat{%
		\label{fig:S12_N1520}
		\includegraphics[clip,width=0.8\columnwidth]{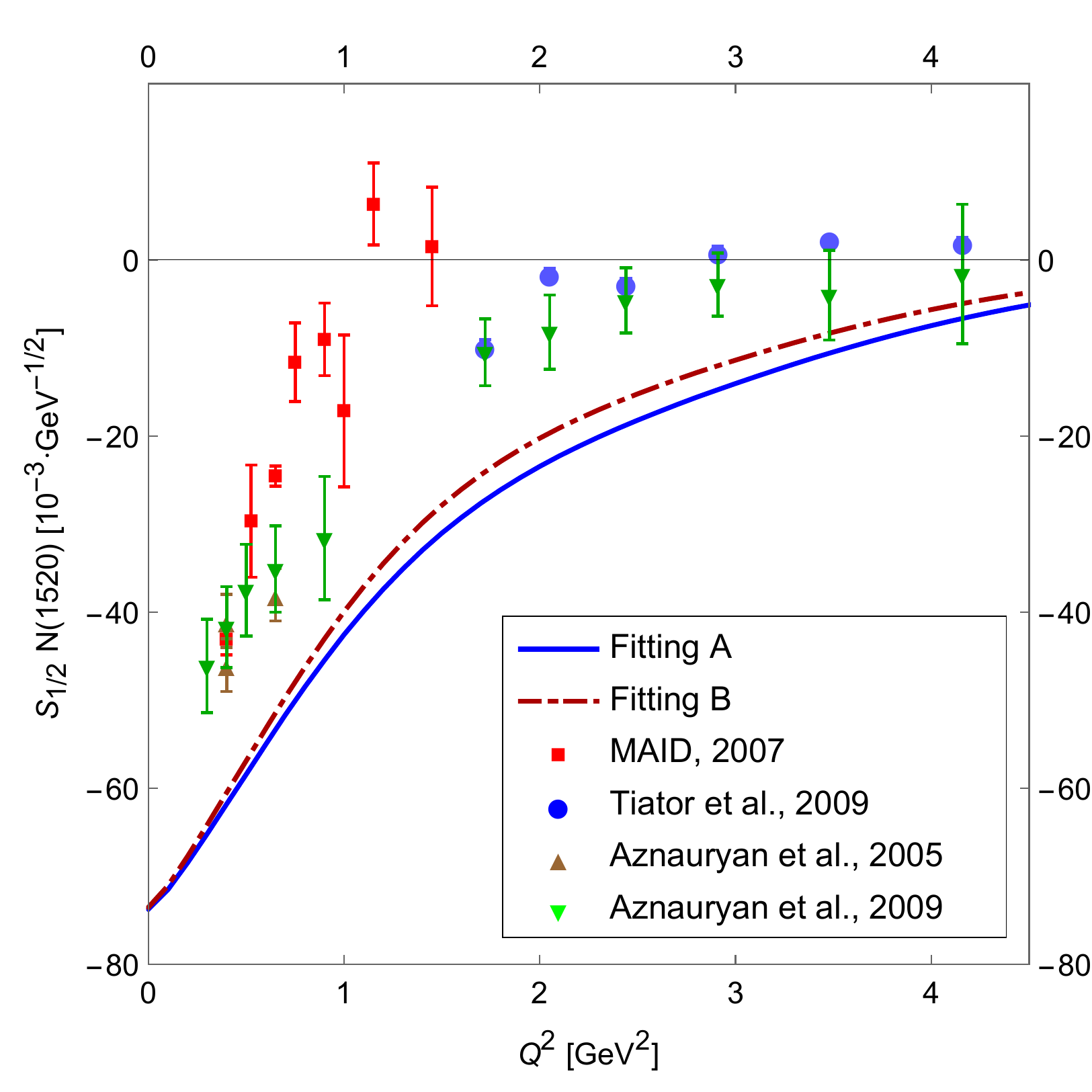}%
	}
	\llap{
	\parbox[b]{4.5in}{\textcolor{white}{\large c}\\\rule{0ex}{2.35in}}}
\llap{
	\parbox[b]{1.5in}{{\scriptsize \cite{Drechsel:2007if}}\\\rule{0ex}{0.782in}
}}
\llap{
	\parbox[b]{1.15in}{{\scriptsize \cite{Tiator:2009mt}}\\\rule{0ex}{0.655in}
}}
\llap{
	\parbox[b]{0.84in}{{\scriptsize \cite{PhysRevC.71.015201}}\\\rule{0ex}{0.523in}
}}
\llap{
	\parbox[b]{0.94in}{{\scriptsize \cite{Aznauryan:2009mx}}\\\rule{0ex}{0.395in}
}}
	\vspace{-1\baselineskip}
	\caption{Helicity amplitudes of the $\gamma^* p\rightarrow N(1520)$ transition compared to the measurements. The solid and dash-dotted curves are the results in Fitting A and Fitting B, respectively. The experimental data are taken from \cite{PhysRevC.71.015201,Aznauryan:2009mx,PhysRevC.86.035203,PhysRevC.93.025206,Drechsel:2007if,Tiator:2009mt,Zyla:2020zbs}.}
	\label{fig:N1520}
\end{figure}

We fit the experimental data on the helicity amplitudes $A_{1/2}$, $A_{3/2}$, and $S_{1/2}$ of the $N(1520)$ and $N(1535)$ resonances in the $\gamma^* p\to N^*$ electromagnetic transition in the constituent quark model in the impulse approximation with assuming that the proton and the resonances are in the three-quark configurations.
The proton spatial wave function is imported from the theoretical results of the proton electric form factor \cite{PhysRevD.105.016008}.
The spin, flavor, and color wave functions of the proton and the resonances are characterized by the $SU(2)$, $SU(3)$, and $SU(3)$ symmetries, respectively. The following calculations are performed:

Fitting A: The helicity amplitudes $A_{1/2}$, $A_{3/2}$, and $S_{1/2}$ of the $N(1520)$ and $N(1535)$ resonances are fitted together, where the $N(1520)$ and $N(1535)$ take the same $L=1$ three quark state.

Fitting B: The helicity amplitudes $A_{1/2}$, $A_{3/2}$, and $S_{1/2}$ of the $N(1520)$ resonance are fitted, where the $N(1520)$ is a $L=1$ three quark state.

Fitting C: The helicity amplitudes $A_{1/2}$, and $S_{1/2}$ of the $N(1535)$ resonances are fitted, where the $N(1535)$ is a $L=1$ three quark state. 

\begin{figure}[!t]
	\captionsetup[subfigure]{labelformat=empty}
	\subfloat[]{%
		\label{fig:A12_N1535}
		\includegraphics[clip,width=0.8\columnwidth]{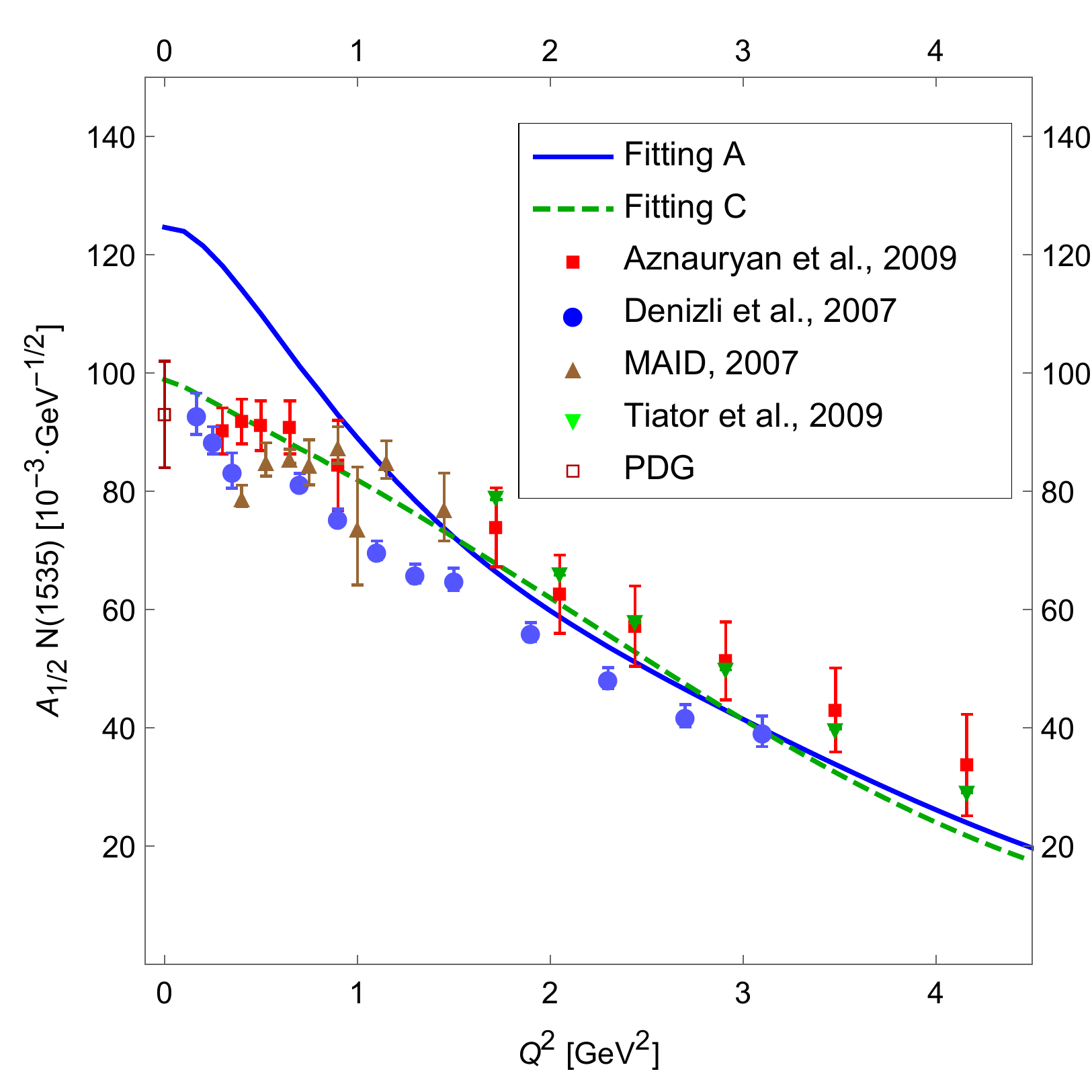}
		\llap{
			\parbox[b]{4.5in}{\textcolor{white}a\\\rule{0ex}{2.3in}
		}}
		\llap{
			\parbox[b]{0.7in}{{\scriptsize \cite{Aznauryan:2009mx}}\\\rule{0ex}{2.01in}
		}}
		\llap{
			\parbox[b]{1in}{{\scriptsize \cite{PhysRevC.76.015204}}\\\rule{0ex}{1.88in}
		}}
		\llap{
			\parbox[b]{1.65in}{{\scriptsize \cite{Drechsel:2007if}}\\\rule{0ex}{1.75in}
		}}
		\llap{
			\parbox[b]{1.3in}{{\scriptsize \cite{Tiator:2009mt}}\\\rule{0ex}{1.62in}
		}}
		\llap{
			\parbox[b]{2.35in}{{\scriptsize \cite{Zyla:2020zbs}}\\\rule{0ex}{1.49in}
		}}
	}
	\vspace{-2.5\baselineskip}
	
	\subfloat[]{%
		\label{fig:S12_N1535}
		\includegraphics[clip,width=0.8\columnwidth]{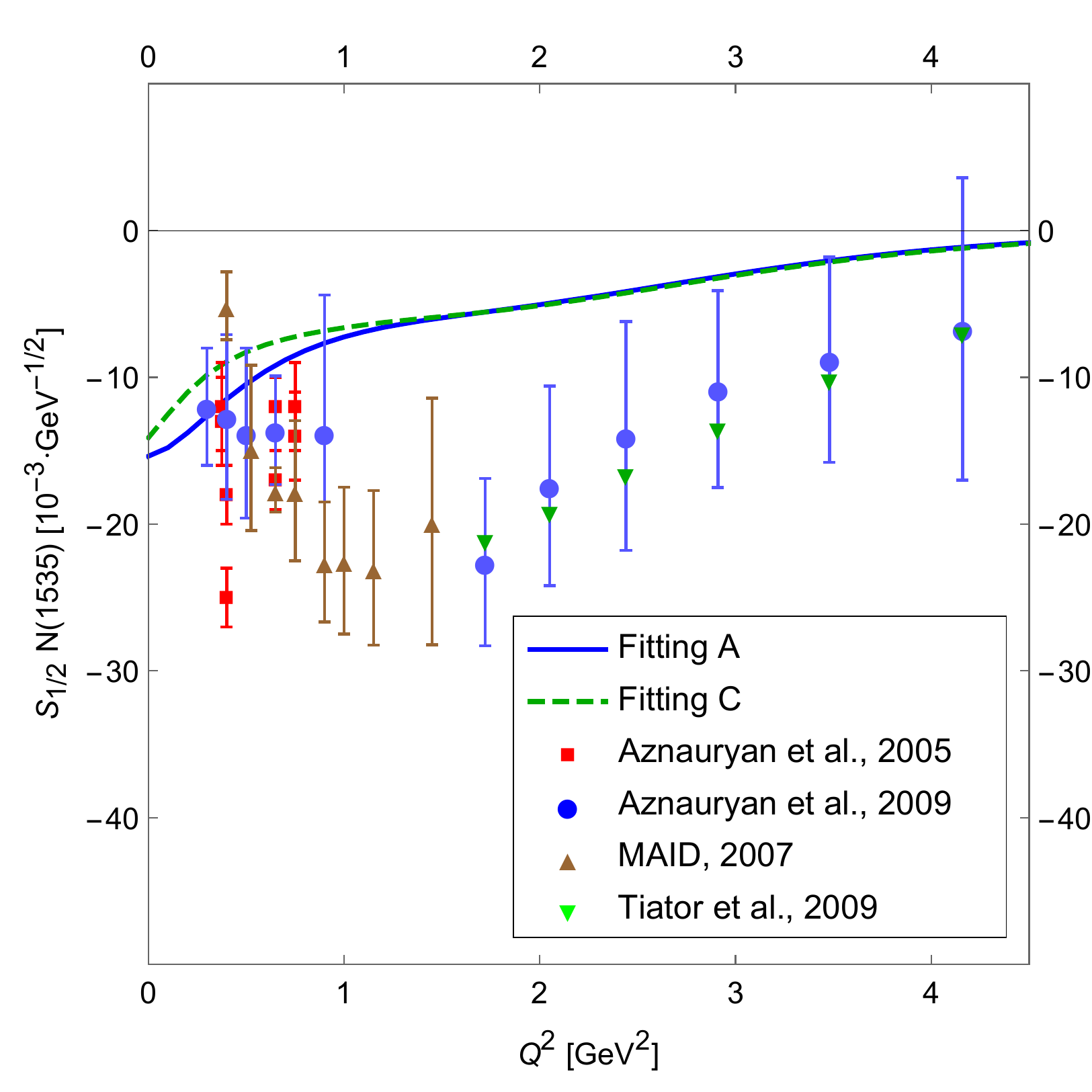}
		\llap{
			\parbox[b]{4.5in}{\textcolor{white}b\\\rule{0ex}{2.3in}
		}}
		\llap{
			\parbox[b]{0.72in}{{\scriptsize \cite{PhysRevC.71.015201}}\\\rule{0ex}{0.785in}
		}}
		\llap{
			\parbox[b]{0.82in}{{\scriptsize \cite{Aznauryan:2009mx}}\\\rule{0ex}{0.655in}
		}}
		\llap{
			\parbox[b]{1.7in}{{\scriptsize \cite{Drechsel:2007if}}\\\rule{0ex}{0.525in}
		}}
		\llap{
			\parbox[b]{1.35in}{{\scriptsize \cite{Tiator:2009mt}}\\\rule{0ex}{0.395in}
		}}
	}
	\vspace{-2\baselineskip}
	\caption{Helicity amplitudes of the $\gamma^* p\rightarrow N(1535)$ transition compared to the measurements. The solid and dashed curves are the results in Fitting A and Fitting C, respectively. The experimental data are taken from \cite{PhysRevC.71.015201,Aznauryan:2009mx,PhysRevC.76.015204,Drechsel:2007if,Tiator:2009mt,Zyla:2020zbs}.}
	\label{fig:N1535}
\end{figure}

The fitting results of the helicity amplitudes of the $N(1520)$ and $N(1535)$ resonances are shown in Figs. \ref{fig:N1520} and \ref{fig:N1535} with the following notations: solid lines indicate Fitting A, dash-dotted lines indicate Fitting B, and dashed lines indicate Fitting C.
The experimental data of the $A_{1/2}$ of the $N(1520)$ resonance clearly split into two clusters, as shown in Fig. \ref{fig:N1520}: the upper one from Ref. \cite{Aznauryan:2009mx} and the lower one from Ref. \cite{Drechsel:2007if,Tiator:2009mt}.
We fit in Fitting A and Fitting B all the $A_{1/2}$ experimental data together.

The experimental data of all helicity amplitudes are reasonably fitted in the three quark picture at large square momentum transfers, $Q^2>2.0$ GeV$^2$ as shown in Figs. \ref{fig:N1520} and \ref{fig:N1535}, suggesting that the two resonances are primarily three quark $L=1$ states. The results are consistent with other quark model calculations.

The fitting results for the amplitudes $A_{3/2}$ and $S_{1/2}$ of the $N(1520)$ resonance, as well as the amplitude $S_{1/2}$ of the $N(1535)$ resonance, are largely discrepant with the experimental data in the low and medium $Q^2$ ranges, which may indicate that the calculations lack the degrees of freedom necessary to fit the experimental data. The $N(1520)$ and $N(1535)$ may contain other components rather than the $L=1$ three-quark state, such as meson clouds, meson-baryon molecules and pentaquarks.

Note that the helicity amplitudes $A_{1/2}$ for both the $N(1520)$ and $N(1535)$ resonances are well fitted over the entire $Q^2$ range. The results contradict the argument that meson-cloud contributions are dominant at low $Q^2$.

It is noted that Fitting B can not significantly improve the results of Fitting A.
It may conclude that the spatial wave functions of the $N(1520)$ and $N(1535)$ resonances are not significantly different in the three-quark picture, which is consistent with mass spectrum calculations, where the tiny $\vec L\cdot \vec S$ coupling results in a very small difference between the two resonances' wave functions.

\section{\label{sec:Summary} Summary}
We have studied the helicity amplitudes of the $N(1520)$ and $N(1535)$ resonances in the process $\gamma^* p\to N^*$.
The impulse approximation is applied in the work, where the electromagnetic current is induced by the free-single-quark currents.

The very similar fitting results in Fitting A, B and C indicate that the quark distributions of the $N(1520)$ and $N(1535)$ resonances in the three quark picture are not significantly different, which is consistent with mass spectrum calculations. 

The experimental data of all the helicity amplitudes are reasonably fitted in the three quark picture at large square momentum transfers, $Q^2>2.0$ GeV$^2$, implying that the $N(1520)$ and $N(1535)$ resonances are dominated by the three valance quarks.

The amplitudes $A_{3/2}$ and $S_{1/2}$ of the $N(1520)$ and the amplitude $S_{1/2}$ of the $N(1535)$ can not be fit in the three quark picture.
It might expect that the $N(1520)$ and $N(1535)$ may consist of other components besides the $L=1$ three quark states, such as meson-clouds, meson-baryon molecules, and pentaquarks. 

It is arguable that meson-cloud contributions are dominant at low $Q^2$, as the helicity amplitudes $A_{1/2}$ for both the $N(1520)$ and $N(1535)$ resonances are reasonably fit in the three quark picture over the entire $Q^2$ range.

Calculations of higher-order contributions from the meson cloud and gluon exchange diagrams, as well as pentaquark components, are currently underway.

\section*{ACKNOWLEDGMENTS}
This research has received funding support from the NSRF via the Program Management Unit for Human Resources \& Institutional Development, Research and Innovation [grant number B05F640055]. A.K. and Y.Y. gratefully acknowledge funding from TSRI-Royal Golden Jubilee Ph.D. (RGJ-PHD) Program (Grant No. PHD/0242/2558) and SUT-OROG scholarship (contract no. 46/2558).\\
%This research was funded in part by (i) Suranaree University of Technology (SUT), (ii) Thailand Science Research Research and Innovation (TSRI), and (iii) National Science Research and Innovation Fund (NSRF) via the Program Management Unit for Human Resources and Institutional Development, Research and Innovation (grant number B05F640055). A.K. and Y.Y. gratefully acknowledge funding from TSRI-Royal Golden Jubilee Ph.D. (RGJ-PHD) Program (Grant No. PHD/0242/2558) and SUT-OROG scholarship program (contract no. 46/2558).\\
	
\begin{appendix}
\section{Matrix elements of $\gamma q\to q'$ for helicity amplitudes}\label{app:T helicity}
The matrix elements of the single quark transition $\gamma q\to q'$ for the helicity $\lambda=0,1$ are derived in detail,
\begin{alignat}{5}\label{eq:T}
&T^0_{\uparrow\uparrow}&&=&&\ \left[ \dfrac{(E'+m)(E+m)}{4E'E}\right]^\frac{1}{2}\left[ 1+ \dfrac{p_z'p_z+2p_-'p_+}{(E'+m)(E+m)}\right], \nonumber\\
&T^0_{\uparrow\downarrow}&&=&&\ \left[ \dfrac{(E'+m)(E+m)}{4E'E}\right]^\frac{1}{2}\left[  \dfrac{\sqrt{2}(p_z'p_--p_-'p_z)}{(E'+m)(E+m)}\right], \nonumber\\
&T^0_{\downarrow\uparrow}&&=&&\ \left[ \dfrac{(E'+m)(E+m)}{4E'E}\right]^\frac{1}{2}\left[  \dfrac{\sqrt{2}(-p_z'p_++p_+'p_z)}{(E'+m)(E+m)}\right], \nonumber\\
&T^0_{\downarrow\downarrow}&&=&&\ \left[ \dfrac{(E'+m)(E+m)}{4E'E}\right]^\frac{1}{2}\left[ 1+\dfrac{p_z'p_z+2p_+'p_-}{(E'+m)(E+m)}\right], \nonumber\\
&T^+_{\uparrow\uparrow}&&=&&\ \left[ \dfrac{(E'+m)(E+m)}{4E'E}\right]^\frac{1}{2}\left[ \dfrac{2p_+}{E+m}\right], \nonumber\\
&T^+_{\uparrow\downarrow}&&=&&\ \left[ \dfrac{(E'+m)(E+m)}{4E'E}\right]^\frac{1}{2}\left[ -\dfrac{\sqrt{2}p_z}{E+m}+\dfrac{\sqrt{2}p_z'}{E'+m}\right], \nonumber\\
&T^+_{\downarrow\uparrow}&&=&&\ 0,\nonumber\\
&T^+_{\downarrow\downarrow}&&=&&\ \left[ \dfrac{(E'+m)(E+m)}{4E'E}\right]^\frac{1}{2}\left[ \dfrac{2p_+'}{E'+m}\right],
\end{alignat}
where $E$ and $E'$ are respectively the energies of the initial and final interacting quarks with the dynamical quark mass of $u$ and $d$ quarks as $m$, and $p_{\pm}=\frac{1}{\sqrt{2}}(p_{x}\pm ip_{y})$.

\end{appendix}
	
\bibliographystyle{unsrtnat}
\bibliography{bibtex}

\begin{thebibliography}{32}
\providecommand{\natexlab}[1]{#1}
\providecommand{\url}[1]{\texttt{#1}}
\expandafter\ifx\csname urlstyle\endcsname\relax
  \providecommand{\doi}[1]{doi: #1}\else
  \providecommand{\doi}{doi: \begingroup \urlstyle{rm}\Url}\fi

\bibitem[Thompson et~al.(2001)]{PhysRevLett.86.1702}
R.~Thompson et~al.
\newblock The
  {$\mathit{ep\ensuremath{\rightarrow}\mathit{e}\ensuremath{'}p\ensuremath{\eta}}$}
  reaction at and above the {$S_{11}(1535)$} baryon resonance.
\newblock \emph{Phys. Rev. Lett.}, 86:\penalty0 1702--1706, Feb 2001.
\newblock \doi{10.1103/PhysRevLett.86.1702}.
\newblock URL \url{https://link.aps.org/doi/10.1103/PhysRevLett.86.1702}.

\bibitem[Denizli et~al.(2007)]{PhysRevC.76.015204}
H.~Denizli et~al.
\newblock ${Q}^{2}$ dependence of the ${S}_{11}(1535)$ photocoupling and
  evidence for a $p$-wave resonance in \ensuremath{\eta} electroproduction.
\newblock \emph{Phys. Rev. C}, 76:\penalty0 015204, Jul 2007.
\newblock \doi{10.1103/PhysRevC.76.015204}.
\newblock URL \url{https://link.aps.org/doi/10.1103/PhysRevC.76.015204}.

\bibitem[Dalton et~al.(2009)]{PhysRevC.80.015205}
M.~M. Dalton et~al.
\newblock Electroproduction of $\ensuremath{\eta}$ mesons in the
  ${S}_{11}(1535)$ resonance region at high momentum transfer.
\newblock \emph{Phys. Rev. C}, 80:\penalty0 015205, Jul 2009.
\newblock \doi{10.1103/PhysRevC.80.015205}.
\newblock URL \url{https://link.aps.org/doi/10.1103/PhysRevC.80.015205}.

\bibitem[Aznauryan et~al.(2005)]{PhysRevC.71.015201}
I.~G. Aznauryan et~al.
\newblock Electroexcitation of the ${P}_{33}(1232)$, ${P}_{11}(1440)$,
  ${D}_{13}(1520)$, and ${S}_{11}(1535)$ at ${Q}^{2}=0.4$ and
  $0.65\phantom{\rule{0.3em}{0ex}}(\mathrm{GeV}/c){}^{2}$.
\newblock \emph{Phys. Rev. C}, 71:\penalty0 015201, Jan 2005.
\newblock \doi{10.1103/PhysRevC.71.015201}.
\newblock URL \url{https://link.aps.org/doi/10.1103/PhysRevC.71.015201}.

\bibitem[Aznauryan et~al.(2008)Aznauryan, Burkert, Kim, Park, Adams, Amaryan,
  Ambrozewicz, Anghinolfi, Asryan, Avakian, Bagdasaryan, Baillie, Ball,
  Baltzell, Barrow, et~al.]{PhysRevC.78.045209}
I.~G. Aznauryan, V.~D. Burkert, W.~Kim, K.~Park, G.~Adams, M.~J. Amaryan,
  P.~Ambrozewicz, M.~Anghinolfi, G.~Asryan, H.~Avakian, H.~Bagdasaryan,
  N.~Baillie, J.~P. Ball, N.~A. Baltzell, S.~Barrow, et~al.
\newblock Electroexcitation of the roper resonance for $1.7 < {Q}^{2} <4.5$
  {GeV}$^{2}$ in $ep\to en\pi^+$.
\newblock \emph{Phys. Rev. C}, 78:\penalty0 045209, Oct 2008.
\newblock \doi{10.1103/PhysRevC.78.045209}.
\newblock URL \url{https://link.aps.org/doi/10.1103/PhysRevC.78.045209}.

\bibitem[Aznauryan et~al.(2009)]{Aznauryan:2009mx}
I.~G. Aznauryan et~al.
\newblock {Electroexcitation of nucleon resonances from CLAS data on single
  pion electroproduction}.
\newblock \emph{Phys. Rev. C}, 80:\penalty0 055203, 2009.
\newblock \doi{10.1103/PhysRevC.80.055203}.

\bibitem[Aznauryan(2003)]{PhysRevC.68.065204}
I.~G. Aznauryan.
\newblock Resonance contributions to $\eta$ photoproduction on protons found
  using dispersion relations and an isobar model.
\newblock \emph{Phys. Rev. C}, 68:\penalty0 065204, Dec 2003.
\newblock \doi{10.1103/PhysRevC.68.065204}.
\newblock URL \url{https://link.aps.org/doi/10.1103/PhysRevC.68.065204}.

\bibitem[Mokeev et~al.(2012{\natexlab{a}})Mokeev, Burkert, Elouadrhiri,
  Fedotov, Golovatch, Gothe, Ishkhanov, Isupov, Adhikari, Aghasyan, Anghinolfi,
  Avakian, Baghdasaryan, Ball, Baltzell, et~al.]{PhysRevC.86.035203}
V.~I. Mokeev, V.~D. Burkert, L.~Elouadrhiri, G.~V. Fedotov, E.~N. Golovatch,
  R.~W. Gothe, B.~S. Ishkhanov, E.~L. Isupov, K.~P. Adhikari, M.~Aghasyan,
  M.~Anghinolfi, H.~Avakian, H.~Baghdasaryan, J.~Ball, N.~A. Baltzell, et~al.
\newblock Experimental study of the ${P}_{11}(1440)$ and ${D}_{13}(1520)$
  resonances from the clas data on
  $ep\ensuremath{\rightarrow}{e}^{\ensuremath{'}}{\ensuremath{\pi}}^{+}{\ensuremath{\pi}}^{\ensuremath{-}}{p}^{\ensuremath{'}}$.
\newblock \emph{Phys. Rev. C}, 86:\penalty0 035203, Sep 2012{\natexlab{a}}.
\newblock \doi{10.1103/PhysRevC.86.035203}.
\newblock URL \url{https://link.aps.org/doi/10.1103/PhysRevC.86.035203}.

\bibitem[Mokeev et~al.(2012{\natexlab{b}})]{Mokeev:2012vsa}
V.~I. Mokeev et~al.
\newblock {Experimental Study of the $P_{11}(1440)$ and $D_{13}(1520)$
  resonances from CLAS data on $ep \rightarrow e'\pi^{+} \pi^{-} p'$}.
\newblock \emph{Phys. Rev. C}, 86:\penalty0 035203, 2012{\natexlab{b}}.
\newblock \doi{10.1103/PhysRevC.86.035203}.

\bibitem[Drechsel et~al.(2007{\natexlab{a}})Drechsel, Kamalov, and
  Tiator]{Drechsel:2007if}
D.~Drechsel, S.~S. Kamalov, and L.~Tiator.
\newblock {Unitary Isobar Model - MAID2007}.
\newblock \emph{Eur. Phys. J. A}, 34:\penalty0 69--97, 2007{\natexlab{a}}.
\newblock \doi{10.1140/epja/i2007-10490-6}.

\bibitem[Tiator et~al.(2011)Tiator, Drechsel, Kamalov, and
  Vanderhaeghen]{Tiator:2011pw}
L.~Tiator, D.~Drechsel, S.~S. Kamalov, and M.~Vanderhaeghen.
\newblock {Electromagnetic Excitation of Nucleon Resonances}.
\newblock \emph{Eur. Phys. J. ST}, 198:\penalty0 141--170, 2011.
\newblock \doi{10.1140/epjst/e2011-01488-9}.

\bibitem[Tiator et~al.(2009)Tiator, Drechsel, Kamalov, and
  Vanderhaeghen]{Tiator:2009mt}
L.~Tiator, D.~Drechsel, S.~S. Kamalov, and M.~Vanderhaeghen.
\newblock {Baryon Resonance Analysis from MAID}.
\newblock \emph{Chin. Phys. C}, 33:\penalty0 1069--1076, 2009.
\newblock \doi{10.1088/1674-1137/33/12/005}.

\bibitem[Capstick and Keister(1995)]{Capstick:1994ne}
Simon Capstick and B.~D. Keister.
\newblock {Baryon current matrix elements in a light front framework}.
\newblock \emph{Phys. Rev. D}, 51:\penalty0 3598--3612, 1995.
\newblock \doi{10.1103/PhysRevD.51.3598}.

\bibitem[Pace et~al.(2000)Pace, Salmè, Cardarelli, and Simula]{PACE200033}
E~Pace, G~Salmè, F~Cardarelli, and S~Simula.
\newblock A light-front description of electromagnetic form factors for {$J \le
  32$} hadrons.
\newblock \emph{Nuclear Physics A}, 666-667:\penalty0 33--37, 2000.
\newblock ISSN 0375-9474.
\newblock \doi{https://doi.org/10.1016/S0375-9474(00)00007-5}.
\newblock URL
  \url{https://www.sciencedirect.com/science/article/pii/S0375947400000075}.
\newblock The Structure of the Nucleon.

\bibitem[Aznauryan and Burkert(2012)]{PhysRevC.85.055202}
I.~G. Aznauryan and V.~D. Burkert.
\newblock Nucleon electromagnetic form factors and electroexcitation of
  low-lying nucleon resonances in a light-front relativistic quark model.
\newblock \emph{Phys. Rev. C}, 85:\penalty0 055202, May 2012.
\newblock \doi{10.1103/PhysRevC.85.055202}.
\newblock URL \url{https://link.aps.org/doi/10.1103/PhysRevC.85.055202}.

\bibitem[Aznauryan and Burkert(2017)]{PhysRevC.95.065207}
I.~G. Aznauryan and V.~D. Burkert.
\newblock Electroexcitation of nucleon resonances of the
  $[70,{1}^{\ensuremath{-}}]$ multiplet in a light-front relativistic quark
  model.
\newblock \emph{Phys. Rev. C}, 95:\penalty0 065207, Jun 2017.
\newblock \doi{10.1103/PhysRevC.95.065207}.
\newblock URL \url{https://link.aps.org/doi/10.1103/PhysRevC.95.065207}.

\bibitem[Ramalho(2017)]{PhysRevD.95.054008}
G.~Ramalho.
\newblock Semirelativistic approximation to the
  ${\ensuremath{\gamma}}^{*}n\ensuremath{\rightarrow}n(1520)$ and
  ${\ensuremath{\gamma}}^{*}n\ensuremath{\rightarrow}n(1535)$ transition form
  factors.
\newblock \emph{Phys. Rev. D}, 95:\penalty0 054008, Mar 2017.
\newblock \doi{10.1103/PhysRevD.95.054008}.
\newblock URL \url{https://link.aps.org/doi/10.1103/PhysRevD.95.054008}.

\bibitem[Ramalho and Pe\~na(2017)]{PhysRevD.95.014003}
G.~Ramalho and M.~T. Pe\~na.
\newblock ${\ensuremath{\gamma}}^{*}n\ensuremath{\rightarrow}{N}^{*}(1520)$
  form factors in the timelike regime.
\newblock \emph{Phys. Rev. D}, 95:\penalty0 014003, Jan 2017.
\newblock \doi{10.1103/PhysRevD.95.014003}.
\newblock URL \url{https://link.aps.org/doi/10.1103/PhysRevD.95.014003}.

\bibitem[Ramalho and Tsushima(2011)]{PhysRevD.84.051301}
G.~Ramalho and K.~Tsushima.
\newblock Simple relation between the
  $\ensuremath{\gamma}n\ensuremath{\rightarrow}n(1535)$ helicity amplitudes.
\newblock \emph{Phys. Rev. D}, 84:\penalty0 051301, Sep 2011.
\newblock \doi{10.1103/PhysRevD.84.051301}.
\newblock URL \url{https://link.aps.org/doi/10.1103/PhysRevD.84.051301}.

\bibitem[Ramalho et~al.(2012)Ramalho, Jido, and Tsushima]{PhysRevD.85.093014}
G.~Ramalho, D.~Jido, and K.~Tsushima.
\newblock Valence quark and meson cloud contributions for the
  $\ensuremath{\gamma}*\ensuremath{\Lambda}\ensuremath{\rightarrow}\ensuremath{\Lambda}*$
  and
  $\ensuremath{\gamma}*{\ensuremath{\Sigma}}^{0}\ensuremath{\rightarrow}\ensuremath{\Lambda}*$
  reactions.
\newblock \emph{Phys. Rev. D}, 85:\penalty0 093014, May 2012.
\newblock \doi{10.1103/PhysRevD.85.093014}.
\newblock URL \url{https://link.aps.org/doi/10.1103/PhysRevD.85.093014}.

\bibitem[Jido et~al.(2008)Jido, D\"oring, and Oset]{PhysRevC.77.065207}
D.~Jido, M.~D\"oring, and E.~Oset.
\newblock Transition form factors of the ${N}^{*}(1535)$ as a dynamically
  generated resonance.
\newblock \emph{Phys. Rev. C}, 77:\penalty0 065207, Jun 2008.
\newblock \doi{10.1103/PhysRevC.77.065207}.
\newblock URL \url{https://link.aps.org/doi/10.1103/PhysRevC.77.065207}.

\bibitem[Golli and \v{S}irca(2013)]{Golli:2013uha}
B.~Golli and S.~\v{S}irca.
\newblock {A chiral quark model for meson electroproduction in the region of
  D-wave resonances}.
\newblock \emph{Eur. Phys. J. A}, 49:\penalty0 111, 2013.
\newblock \doi{10.1140/epja/i2013-13111-y}.

\bibitem[Julia-Diaz et~al.(2008)Julia-Diaz, Lee, Matsuyama, Sato, and
  Smith]{JuliaDiaz:2007fa}
B.~Julia-Diaz, T.~S.~H. Lee, A.~Matsuyama, T.~Sato, and L.~C. Smith.
\newblock {Dynamical coupled-channels effects on pion photoproduction}.
\newblock \emph{Phys. Rev. C}, 77:\penalty0 045205, 2008.
\newblock \doi{10.1103/PhysRevC.77.045205}.

\bibitem[An and Zou(2009)]{An2009}
C.~S. An and B.~S. Zou.
\newblock The role of the $qqqq\bar{q}$ components in the electromagnetic
  transition {$\gamma^*N\rightarrow N^*(1535)$}.
\newblock \emph{The European Physical Journal A}, 39\penalty0 (2):\penalty0
  195--204, Feb 2009.
\newblock ISSN 1434-601X.
\newblock \doi{10.1140/epja/i2008-10698-x}.
\newblock URL \url{https://doi.org/10.1140/epja/i2008-10698-x}.

\bibitem[Xu et~al.(2020)Xu, Kaewsnod, Zhao, Liu, Srisuphaphon, Limphirat, and
  Yan]{PhysRevD.101.076025}
K.~Xu, A.~Kaewsnod, Z.~Zhao, X.~Y. Liu, S.~Srisuphaphon, A.~Limphirat, and
  Y.~Yan.
\newblock Pentaquark components in low-lying baryon resonances.
\newblock \emph{Phys. Rev. D}, 101:\penalty0 076025, Apr 2020.
\newblock \doi{10.1103/PhysRevD.101.076025}.
\newblock URL \url{https://link.aps.org/doi/10.1103/PhysRevD.101.076025}.

\bibitem[Kaewsnod et~al.(2022)]{PhysRevD.105.016008}
A.~Kaewsnod et~al.
\newblock Study of $n(1440)$ structure via
  ${\ensuremath{\gamma}}^{*}p\ensuremath{\rightarrow}n(1440)$ transition.
\newblock \emph{Phys. Rev. D}, 105:\penalty0 016008, Jan 2022.
\newblock \doi{10.1103/PhysRevD.105.016008}.
\newblock URL \url{https://link.aps.org/doi/10.1103/PhysRevD.105.016008}.

\bibitem[Xu et~al.(2019)Xu, Kaewsnod, Liu, Srisuphaphon, Limphirat, and
  Yan]{PhysRevC.100.065207}
K.~Xu, A.~Kaewsnod, X.~Y. Liu, S.~Srisuphaphon, A.~Limphirat, and Y.~Yan.
\newblock Complete basis for the pentaquark wave function in a group theory
  approach.
\newblock \emph{Phys. Rev. C}, 100:\penalty0 065207, Dec 2019.
\newblock \doi{10.1103/PhysRevC.100.065207}.
\newblock URL \url{https://link.aps.org/doi/10.1103/PhysRevC.100.065207}.

\bibitem[Dreschsel and Tiator(1992)]{Dreschsel_1992}
D~Dreschsel and L~Tiator.
\newblock Threshold pion photoproduction on nucleons.
\newblock \emph{Journal of Physics G: Nuclear and Particle Physics},
  18\penalty0 (3):\penalty0 449--497, mar 1992.
\newblock \doi{10.1088/0954-3899/18/3/004}.
\newblock URL \url{https://doi.org/10.1088/0954-3899/18/3/004}.

\bibitem[Drechsel et~al.(1999)Drechsel, Hanstein, Kamalov, and
  Tiator]{DRECHSEL1999145}
D.~Drechsel, O.~Hanstein, S.S. Kamalov, and L.~Tiator.
\newblock A unitary isobar model for pion photo- and electroproduction on the
  proton up to 1 gev.
\newblock \emph{Nuclear Physics A}, 645\penalty0 (1):\penalty0 145--174, 1999.
\newblock ISSN 0375-9474.
\newblock \doi{https://doi.org/10.1016/S0375-9474(98)00572-7}.
\newblock URL
  \url{https://www.sciencedirect.com/science/article/pii/S0375947498005727}.

\bibitem[Drechsel et~al.(2007{\natexlab{b}})Drechsel, Kamalov, and
  Tiator]{Drechsel2007}
D.~Drechsel, S.~S. Kamalov, and L.~Tiator.
\newblock Unitary isobar model --maid2007.
\newblock \emph{The European Physical Journal A}, 34\penalty0 (1):\penalty0 69,
  Oct 2007{\natexlab{b}}.
\newblock ISSN 1434-601X.
\newblock \doi{10.1140/epja/i2007-10490-6}.
\newblock URL \url{https://doi.org/10.1140/epja/i2007-10490-6}.

\bibitem[Mokeev et~al.(2016)Mokeev, Burkert, Carman, Elouadrhiri, Fedotov,
  Golovatch, Gothe, Hicks, Ishkhanov, Isupov, and
  Skorodumina]{PhysRevC.93.025206}
V.~I. Mokeev, V.~D. Burkert, D.~S. Carman, L.~Elouadrhiri, G.~V. Fedotov, E.~N.
  Golovatch, R.~W. Gothe, K.~Hicks, B.~S. Ishkhanov, E.~L. Isupov, and Iu.
  Skorodumina.
\newblock New results from the studies of the $n(1440)1/{2}^{+},
  n(1520)3/{2}^{\ensuremath{-}}$, and
  $\mathrm{\ensuremath{\Delta}}(1620)1/{2}^{\ensuremath{-}}$ resonances in
  exclusive
  $ep\ensuremath{\rightarrow}{e}^{\ensuremath{'}}{p}^{\ensuremath{'}}{\ensuremath{\pi}}^{+}{\ensuremath{\pi}}^{\ensuremath{-}}$
  electroproduction with the clas detector.
\newblock \emph{Phys. Rev. C}, 93:\penalty0 025206, Feb 2016.
\newblock \doi{10.1103/PhysRevC.93.025206}.
\newblock URL \url{https://link.aps.org/doi/10.1103/PhysRevC.93.025206}.

\bibitem[Zyla et~al.(2020)]{Zyla:2020zbs}
P.A. Zyla et~al.
\newblock {Review of Particle Physics}.
\newblock \emph{PTEP}, 2020\penalty0 (8):\penalty0 083C01, 2020.
\newblock \doi{10.1093/ptep/ptaa104}.

\end{thebibliography}
	
\end{document}